\documentclass[uplatex, aip, jap, cp, preprent, graphicx, onecolumn]{revtex4-1}

\usepackage{graphicx}
\usepackage{dcolumn}
\usepackage{bm}
\usepackage{here}
\usepackage{amsmath,amssymb}
\usepackage{braket}
\graphicspath{{fig/}{../fig/}}

\newcommand{\chem}[1]{$\mathrm{#1}$}

\renewcommand{\Im}{\mathrm{Im}}

\renewcommand{\i}{\mathrm{i}}
\newcommand{\Tr}{\mathrm{Tr}}

\newcommand{\Pauli}{\hat{\bm{\sigma}}}

\newcommand{\EF}{E_{\mathrm{F}}}
\newcommand{\kF}{k_\mathrm{F}}

\newcommand{\cd}{c^{\dagger}}
\newcommand{\dd}{d^{\dagger}}

\newcommand{\ck}{c_{\bm{k}}}
\newcommand{\cdk}{\cd_{\bm{k}}}

\newcommand{\Hs}{H_\mathrm{s}}

\newcommand{\Himp}{H_\mathrm{imp}}

\newcommand{\Hhyb}{H_\mathrm{hyb}}
\newcommand{\Hcf}{H_\mathrm{cf}}
\newcommand{\Hso}{H_{\mathrm{so}}}

\newcommand{\Vsd}{V^\mathrm{sd}}

\newcommand{\Hex}{H_{\mathrm{ex}}}

\newcommand{\Gs}{G_{\mathrm{s}}}
\newcommand{\GsTilde}{\tilde{G}_{\mathrm{s}}}

\newcommand{\Gimp}{G_{\mathrm{imp}}}

\newcommand{\Dels}{\Delta_\mathrm{s}}
\newcommand{\Deld}{\Delta_\mathrm{d}}

\newcommand{\Delc}{\Delta_\mathrm{cf}}

\newcommand{\Nimp}{N_{\mathrm{imp}}}
\newcommand{\nimp}{n_{\mathrm{imp}}}
\newcommand{\dwidth}{\varGamma_{0}}

\newcommand{\rup}{\rho_{+}}
\newcommand{\rdn}{\rho_{-}}

\newcommand{\dgxy}{x^{2}-y^{2}}
\newcommand{\dgzr}{3z^{2}-r^{2}}
\newcommand{\rate}{\alpha}
\newcommand{\rhoss}{\rho^{\mathrm{ss}}}
\newcommand{\rhosd}{\rho^{\mathrm{sd}}}

\newcommand{\coef}{w}
\newcommand{\AMR}{\mathrm{AMR}}
\newcommand{\selfenergyss}{\eta^{\mathrm{ss}}}
\newcommand{\selfenergysd}{\varSigma^{\mathrm{sd},\pm}}

\begin{document}
\title[Theoretical Study on Four-fold Symmetric Anisotropic Magnetoresistance Effect in Cubic Single-crystal Ferromagnetic Model]{Theoretical Study on Four-fold Symmetric Anisotropic Magnetoresistance Effect in Cubic Single-crystal Ferromagnetic Model}
\author{Y. Yahagi}
\altaffiliation{	Applied physics, Tohoku University, Sendai, Miyagi, Japan.}
\email{yahagi@solid.apph.tohoku.ac.jp}
\author{D. Miura}
\author{A. Sakuma}
\affiliation{Tohoku Univ.}
\date{\today}

\begin{abstract}
	In this study, we present a theoretical interpretation of the experimental results that the anisotropic magnetoresistance (AMR) effect has a four-fold symmetric component, $c_4$, in cubic ferromagnetic metals. 
	The theoretical model that we employ is based on the Anderson impurity model that includes a four-fold symmetric crystalline electric field, and we assume that the impurities have 3d electron orbitals and spin--orbit interaction (SOI). We describe the DC conductivity on the basis of the Kubo formula, and we investigate $c_4$ by analyzing the magnetization direction dependence of the resultant AMR ratio. 
	Analytical and numerical calculations are performed; 
	the analytical calculation reveals that $c_4$ arises from the fourth-order contribution of the SOI, and the numerical calculation provides the parameter dependencies of $c_{4}$ in our model.
	From the calculation results, we observe that the splitting of impurity 3d levels due to SOI is responsible for the existence of $c_{4}$ in cubic ferromagnetic metals.

\end{abstract}

\maketitle

\section{Introduction}
 The anisotropic magnetoresistance (AMR) effect has been known to be a special magnetoresistance effect that occurs in ferromagnetic metals and has had applications in magnetic sensors.
 Owing the theoretical research conducted in the 1960s and 1970s, it is widely believed that this is a spin-dependent transport property of ferromagnets. 
 The efficiency of AMR is referred using the AMR ratio defined as
 \begin{equation} \label{eq:AMRratio}
 \AMR(\phi)\equiv \frac{\Delta \rho(\phi)}{\rho_{\perp}}=\frac{\rho(\phi)-\rho_{\perp}}{\rho_{\perp}},
 \end{equation}
with $\rho_{\perp}=\rho(\frac{\pi}{2})$.
Usually, in experiments on bulk polycrystalline ferromagnetic metals, 
the angular dependence of resistivity is phenomenologically written as
$\AMR(\phi)=c_{0}+c_{2}\cos 2\phi$,
where $c_{n}$ denotes the coefficient of the $\cos n\phi$ component. Theoretically, the AMR effect has been successfully explained using an s-d impurity scattering model considering spin--orbit interactions (SOIs).\cite{Berger1968,Campbell1970,Potter1974,Mcguire1975,Kokado2012}

In recent years, the AMR effect has attracted considerable attention in the field of spintronics 
because it is a type of SOI-related phenomenon that is expected to be a key aspect in controlling the magnetization alignments of multilayer systems using an electric field.
In the line of this research, so-called perpendicular AMR effects were observed in the magnetic multilayer systems 
where the AMR effect depends not only on the relative angle between $\bm{M}$ and $\bm{J}$ but also on the $\bm{M}$ angle measured in the plane perpendicular to $\bm{J}$.
We successfully provided a theoretical description for the effects based on the tight-binding model, including the Rashba-type SOI at the interface.\cite{Yahagi2018}
The mechanism we observed is closely related to the Edelstein effect, which is one of the causes of spin--orbit torque acting on the magnetization at the magnetic multilayer interfaces.

\begin{figure}[H]
	\centering
	\includegraphics[clip, width=0.6\columnwidth]{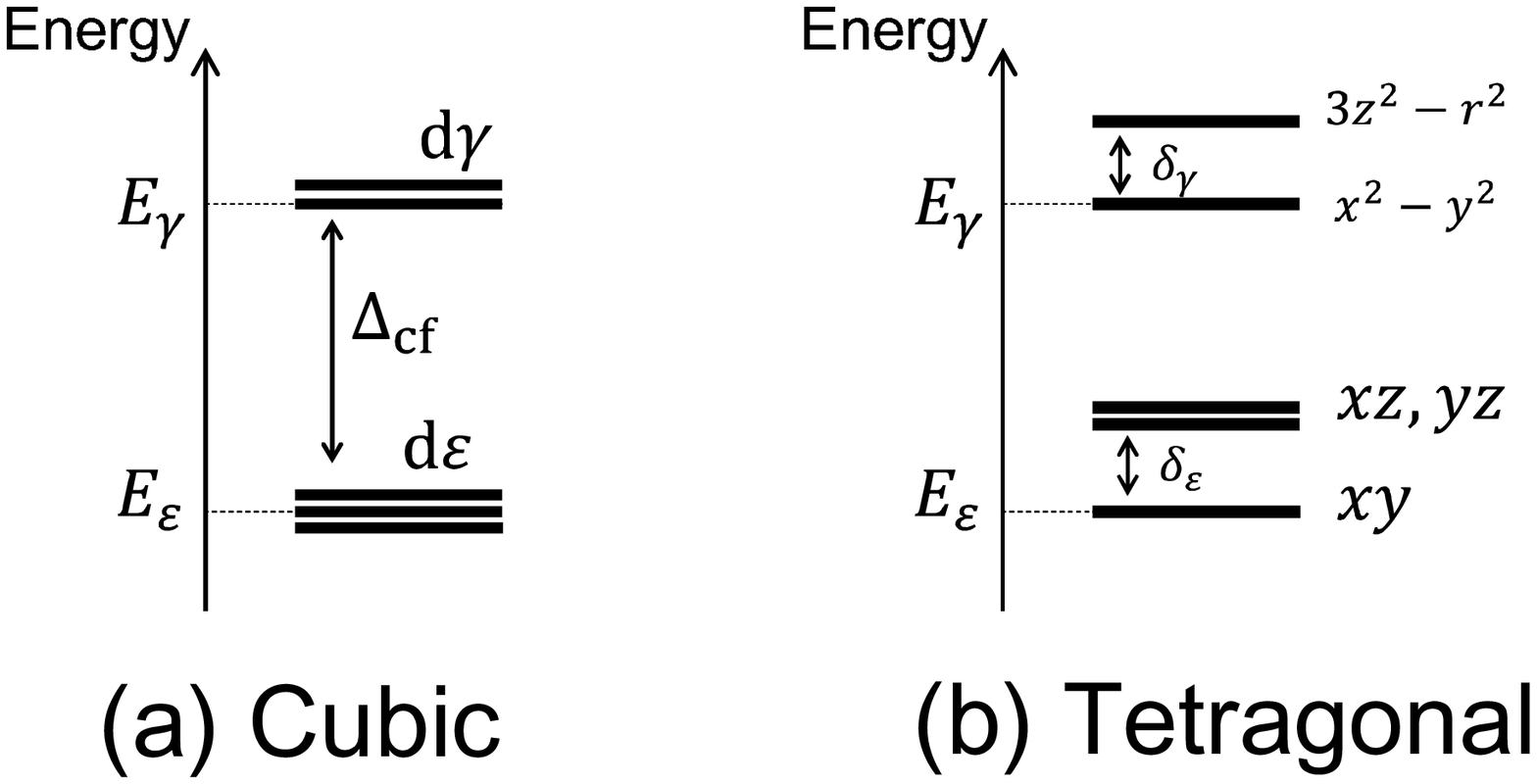}
	\caption{Energy levels of the 3d states in the crystal field of (a) cubic symmetry and (b) tetragonal symmetry.}
	\label{fig:CF}
\end{figure}

Recently, interesting behaviors have been noticed in some single-crystal ferromagnets such as \chem{Fe_{4}N}\cite{Tsunoda2010,Ito2012,Ito2014,Kabara2014} and \chem{Co_{2}MnSi}\cite{Oogane2018}
wherein the AMR ratio exhibits four-fold symmetry in the form of
$
\AMR(\phi)=c_{0}+c_{2}\cos 2\phi+c_{4}\cos 4\phi.
$
In 2015, Kokado and Tsunoda proposed a theory for explaining the origin of four-fold term,
wherein tetragonal symmetric crystal fields 
are responsible for the $c_{4}$ term from the second-order perturbation expansion in terms of SOI.\cite{Kokado2015} 
In their study, they found that $c_{4}$ is proportional to the deference in the projected density of states (PDOS) at the Fermi energy ($\EF$) among tetragonal splitting d$\varepsilon$ states (Fig.\ref{fig:CF}), 
$
c_{4}\propto D^{(d)}_{xy,+} -D^{(d)}_{yz,+},
$
where $D^{(d)}_{m,\sigma}$ represents the PDOS of $m$ state of 3d orbitals ($m=xy,yz,zx,x^{2}-y^{2},3z^{2}-r^{2}$) 
with spin $\sigma=\pm$. \cite{Kokado2015}
The spin of $+(-)$ indicates the majority (minority) spin state whose quantization axis has the same direction as $\hat{M}$.

Assuming planar or uniaxial lattice distortions on films, the above explanation may be applied to account for the presence of $c_{4}$ because such distortions change crystal symmetry from cubic to tetragonal. However, symmetry transitions have not been directly observed even when a finite $c_4$ appears. Therefore, for understanding the four-fold AMR effects, it is worth further focusing on cubic systems.

In this study, we show the presence of not only $c_{2}$ but also $c_{4}$ on cubic single-crystal ferromagnetic 3d alloys. 
Inspired by the Kokado model,\cite{Kokado2015} we use the s-d impurity scattering model with cubic crystal fields and SOI. 
The AMR is treated on the basis of the Kubo--Greenwood formula.
With this approach, we can consider the non-perturbative role of SOIs. 
To clarify the physical aspect of $c_{4}$, we first perform 
perturbative calculations with respect to the SOI.
From the analysis, we observed that the fourth-order term of SOI gives rise to the splitting of d$\varepsilon$ states, resulting in the appearance of $c_{4}$ term.
Next, we show the unperturbative result using numerical calculations to evaluate the behavior of $c_4$ 
as functions of angles $\phi$, SOI-strengths, and $\EF$. Finally, we provide a summary and conclude this study.

\section{MODEL AND FORMULATION}
In this section, we present the model Hamiltonian $H$ and formulation for AMRs.
Assuming a cubic single-crystal ferromagnetic 3d alloy system, AMR is described by using the impurity scattering model. 
Here, the 3d-electrons are relatively localized and are then assumed contribute little to conduction.
Therefore, we regard the 3d-band to act only as a ferromagnetic background. 
In this situation, only the 4s-electrons contribute to conduction and resistivity is governed by s-d impurity scattering. 
We regard the impurity atoms to have a magnetic 3d character.
Thus, we adopt the multi-orbital d-impurity Anderson model to describe the above situation as follows:
\begin{equation}
H=\Hs(\phi) + \Himp(\phi) +\Hhyb,
\end{equation}
where $\Hs(\phi)$ is the 4s-conduction electron Hamiltonian, $\Himp(\phi)$ represents the impurity 3d states, 
and $\Hhyb$ denotes the s-d hybridization term. 
The conduction electrons are treated within the electron-gas model with exchange splitting from a ferromagnetic background sustained by the 3d-bands
\begin{equation}
\Hs(\phi)=\sum_{\bm{k}} (E_{0}+tk^{2}) \cd_{\bm{k}} c_{\bm{k}}
-\Dels  \sum_{\bm{k}} ( \cd_{\bm{k}}\Pauli c_{\bm{k}} ) \cdot \hat{\bm{M}}(\phi),
\label{eq:partial_hamiltonian_sk}
\end{equation}
with
\begin{align}
 \Pauli=(\sigma_{x},\sigma_{y},\sigma_{z}), \\
 \hat{\bm{M}}(\phi) = (\cos\phi,\sin\phi,0),
\end{align}
where $c_{\bm{k}}=(c_{\bm{k},\uparrow},c_{\bm{k},\downarrow})^{\mathrm{T}}$ and $\cd_{\bm{k}}=(\cd_{\bm{k},\uparrow},\cd_{\bm{k},\downarrow})$ 
are the spinor-represented operators creating and annihilating the conduction electron state with the wave vector $\bm{k}$,
and $\sigma_{\mu}$ is the $\mu$-component of the Pauli matrix.
The first term represents the kinetic energy $tk^{2}$ and the bottom energy $E_{0}$,
and $\Dels$ is the strength of the exchange splitting on conduction band. 

Impurity 3d states are treated as localized 3d atomic orbitals with exchange splitting, SOIs, and crystal fields of cubic symmetry
reflecting the 3d host matrix.
\begin{align}
\Himp(\phi)&=\sum_{i}^{N_\mathrm{imp}}\Himp^{(i)}(\phi)\\
\Himp^{(i)}(\phi)&= \Hex^{(i)}(\phi) + \Hcf^{(i)} + \Hso^{(i)}\label{eq:HamiltonianImpurity}.
\end{align}
with
\begin{align}
\Hex^{(i)}(\phi)=&-\Deld\sum_{m} ( \dd_{i,m} \Pauli d_{i,m} )\cdot \hat{\bm{M}}(\phi),  \\
\Hcf^{(i)} = &E_{\varepsilon}(\dd_{i,xy}d_{i,xy}+\dd_{i,yz}d_{i,yz}+\dd_{i,xz}d_{i,xz}) \nonumber \\
&+  E_{\gamma}(\dd_{i,x^{2}-y^{2}}d_{i,x^{2}-y^{2}}+\dd_{i,3z^{2}-r^{2}}d_{i,3z^{2}-r^{2}}), \\
\Hso =& \frac{\lambda}{2} \sum_{m,m'} \dd_{i,m}(\Pauli\cdot\bm{l})_{m,m'}d_{i,m'},
\label{eq:SOI_Hamiltonian}
\end{align}
where the suffix $i$ indicates the site index of the impurity position.
$d_{i,m}=(d_{i,m,\uparrow},d_{i,m,\downarrow})^{\mathrm{T}}$ and $\dd_{i,m}=(\dd_{i,m,\uparrow},\dd_{i,m,\downarrow})$ 
are the spinor-represented operators creating and annihilating the impurity 3d state with the site $i$ and the orbital $m$.
$\Deld$ is the strength of the exchange splitting on impurity states
with the polarization direction of $\hat{\bm{M}}(\phi)$.
$E_{\varepsilon(\gamma)}$ is the energy level of d$\varepsilon$(d$\gamma$) state.
$\bm{l}=(l_{x},l_{y},l_{z})$ is the angular momentum operator of  $l=2$ and $\lambda$ is the coupling constant of SOI.
We note that Eqs.\eqref{eq:HamiltonianImpurity}-\eqref{eq:SOI_Hamiltonian} are in the same form as the model in Kokado’s study.\cite{Kokado2015}

The s-d hybridization between the conduction band and impurity states is written as
\begin{align}
\Hhyb=\sum_{i}^{N_\mathrm{imp}} \sum_{\bm{k},m}\left( \Vsd_{\bm{k};i,m} \cd_{\bm{k}}d_{i,m} + H.c. \right),\\
\Vsd_{\bm{k};i,m} = -\frac{e^{\i \bm{k}\cdot \bm{r}_{i}}}{\sqrt{\Omega}}f(k)X_{2,m}(\theta_{\bm{k}},\phi_{\bm{k}}),
\end{align}
where  $ \bm{r}_{i}$ is the position of impurity center,  $\Omega$ is the volume of the system, $f(k)$ is the isotropic coefficient originating from the radial part of the 3d orbital, and $X_{l,m}(\theta_{\bm{k}},\phi_{\bm{k}})$ is the cubic harmonics in k-space given by
\begin{equation} \label{eq:CubicHarmonic}
X_{2,m}(\theta_{\bm{k}},\phi_{\bm{k}}) = 
\begin{cases}
\sqrt{\frac{5}{4\pi}}\frac{1}{2}(\cos^{2}\theta_{\bm{k}}-1)&m=\dgzr, \\
\sqrt{\frac{5}{4\pi}}\frac{\sqrt{3}}{2}\sin^{2}\theta_{\bm{k}}(\cos^{2}\phi_{\bm{k}}-\sin^{2}\phi_{\bm{k}})&m=\dgxy,\\
\sqrt{\frac{5}{4\pi}}\sqrt{3}\cos\theta_{\bm{k}}\sin\theta_{\bm{k}}\cos\phi_{\bm{k}}&m=zx,\\
\sqrt{\frac{5}{4\pi}}\sqrt{3}\cos\theta_{\bm{k}}\sin\theta_{\bm{k}}\sin\phi_{\bm{k}}&m=yz, \\
\sqrt{\frac{5}{4\pi}}\sqrt{3}\sin^{2}\theta_{\bm{k}}\cos\phi_{\bm{k}}\sin\phi_{\bm{k}}&m=xy.
\end{cases}
\end{equation}
The difference between our model and Kokado \textit{et al.}’s model lies in that the treatment of s-d impurity scattering 
where the conductive s-electrons are scattered into the impurity d-states in our model 
while the s-electrons are scattered into host d-bands through the impurity atoms in their model.
As the polarization direction and the spatial symmetry of d-charactor are taken into account in both cases in the scattering events,
both models essentially provide the same picture in terms of AMR symmetry.

To describe AMR, we investigate the conductivity changes of the system in a microscopic manner. 
The longitudinal conductivity $\sigma_{xx}(\phi)(=1/\rho(\phi))$ at zero temperature is given by the Kubo--Greenwood formula\cite{Kubo1957,Greenwood1958}:
\begin{equation} \label{eq:Kubo-Greenwood_0K}
\sigma_{xx}(\phi)=-\frac{\hbar}{4\pi \Omega}\Tr_{\bm{k},\sigma}\braket{J_{x}\{\Gs^+(\phi)-\Gs^-(\phi)\}J_{x}\{\Gs^+(\phi)-\Gs^-(\phi)\}}_\mathrm{conf},
\end{equation}
where $\Gs^{\pm}(\phi)$ is the conduction electron's Green's function ($+$: retarded, $-$: advanced) at the Fermi  level $\EF$
and $\braket{\cdots}_{\mathrm{conf}}$ indicates the configuration average for impurities. 
The charge current operators $J_{x}$ are expressed by
\begin{equation}\label{eq:CurrentOperator}
J_{x}=-\frac{2te}{\hbar}\sum_{\bm{k}}k_{x} \cdk\ck,
\end{equation}
where $e$ denotes the elementary charge.
The anomalous currents from the s-d hybridizations are neglected because its contribution seems much smaller than that from normal currents.
To perform practical calculations, we employ the first Born approximation, and  $\Gs^{\pm}(\phi)$ is replaced by impurity averaged Green's function: 
\begin{equation}
\GsTilde^{\pm}(\phi) = (\EF-\Hs(\phi)-\varSigma^{\pm}(\phi))^{-1},
\end{equation}
\begin{equation}
\varSigma^{\pm}(\phi)\equiv \pm\i \selfenergyss + \selfenergysd(\phi),
\end{equation}
where $\selfenergyss$ is a positive parameter representing the self-energy from the s-s scattering and $\selfenergysd(\phi)$ is the self-energy from the s-d scattering.
 $\selfenergysd(\phi)$ is written as
\begin{equation}\label{eq:selfenergy}
\selfenergysd(\phi)=\nimp |f(\kF)|^{2} \sum_{\bm{k},m,m',\sigma,\sigma'} X^{\ast}_{2,m'}(\theta_{\bm{k}},\phi_{\bm{k}})X_{2,m}(\theta_{\bm{k}},\phi_{\bm{k}})
\braket{[\Gimp(\phi)]_{i,m,\sigma;i,m',\sigma'}}_{\mathrm{conf}},
\end{equation}
where $\nimp\equiv\Nimp / \Omega$, and $\Gimp^{\pm}(\phi)$ is the impurity Green's function at $\EF$,
\begin{equation}\label{eq:impGreenfunc}
\Gimp^{\pm}(\phi)=(\EF-\Himp(\phi)\pm\i \dwidth)^{-1}.
\end{equation}
In this expression, the finite energy width $\dwidth$ is phenomenologically introduced as reflecting the hybridization with the host 3d bands.
As we consider the random impurities,  
the variables in Eqs. \eqref{eq:selfenergy} and \eqref{eq:impGreenfunc} do not depend on the impurity site $i$;
Hereinafter, the suffix $i$ is omitted.
Then, Eq. \eqref{eq:Kubo-Greenwood_0K} is rewritten as
\begin{equation}\label{eq:KG_2}
\sigma_{xx}(\phi)
=-\frac{\hbar}{4\pi\Omega} \Tr_{\bm{k},\sigma} J_{x}\{\GsTilde^+(\phi)-\GsTilde^-(\phi)\}J_{x}\{\GsTilde^+(\phi)-\GsTilde^-(\phi)\}.
\end{equation}
In this case, there is no contribution from vertex correction because $\delta \varSigma / \delta \Gs = 0$.

\section{Results and Discussion}
We perform analytical calculations to extract a mechanism
and the numerical calculations to see the detailed trend of AMR on cubic symmetry.
\subsection{Perturbative analysis in lifetime approximation} 
We herein show the results that finite $c_4$ can be obtained from the fourth-order perturbation with respect to the SOI.
For the analytical calculations, we first take the following three approximations: (1) two-current model as
\begin{equation}\label{eq:Two_current}
\rho(\phi) \simeq\left(\frac{1}{\rup(\phi)}+\frac{1}{\rdn(\phi)}\right)^{-1},
\end{equation}
where $\rup(\rdn)$ indicates the majority (minority) spin resistor; 
(2) Matthiessen's rule as
\begin{equation}
\label{eq:Matthiessen}
\rho_{\sigma}(\phi) \simeq \rhoss_{\sigma} + \rhosd_{\sigma}(\phi),
\end{equation}
where $\rhoss (\rhosd)$ is the resistivity originating from s-s (s-d) scattering; 
(3) Lifetime approximation for s-d scattering as
\begin{align}
\rhosd_{\sigma}(\phi) \simeq \frac{m^{\ast}}{e^{2}n_{\sigma}}\left(-\frac{1}{\hbar}\Im\varSigma^{+}_{\sigma,\sigma}(\phi) \right),
\end{align} 
where $m^{\ast}$ denotes the effective mass of electron
and $n_{\sigma}$ is the electron concentration of $\sigma$ spin at $\EF$.
 Incidentally, $\rhoss_{\sigma}$ is treated as the constant parameter. 
 \\

We next take the higher-order perturbation expansion with respect to the SOI in $\Gimp^{+}$:
\begin{equation}\label{eq:T-matrix}
\Gimp^{+}(\phi) = (1+gT(\phi))g
\end{equation}
where $T(\phi)\equiv  \Hso +  \Hso g \Hso +\cdots$ is T-matrix and $g(\phi)=\{\EF-(\Hex^{(i)}(\phi)+\Hcf^{(i)})-\i\dwidth \}^{-1}$ is unperturbed Green's function of the impurity state.
The argument $\phi$ is omitted from $g(\phi)$ owing to the paper savings.
In the cubic system, the relations of $g_{xy, \sigma}=g_{yz, \sigma}=g_{zx,\sigma}\equiv g_{\varepsilon, \sigma}$ and 
$g_{x^{2}-y^{2}, \sigma}=g_{3z^{2}-r^{2}, \sigma} \equiv g_{\gamma, \sigma}$ are satisfied.

Here, we have $\rhosd_{\sigma}(\phi)=\rhosd_{0,\sigma}+\rho'_{\sigma}(\phi)$ 
where $\rhosd_{0,\sigma}$ and $\rho'_{\sigma}(\phi)$ are the unperturbed term and the perturbed term is
($\rhosd_{0,\sigma} \gg \rho'_{\sigma}(\phi)$). 
Then, $\rho(\phi)$ is derived as
\begin{equation} \label{eq:resistivity1}
\rho(\phi)\simeq \rho_{0}+ \frac{\rho'_{+}(\phi)+\rate^{2}\rho'_{-}(\phi)}{(1+\alpha)^{2}},
\end{equation}
with $\rho_{0}\equiv (\rho_{0,+}^{-1}+\rho_{0,-}^{-1})^{-1}$, $\rho_{0,\sigma}\equiv\rhoss_{\sigma}+\rhosd_{0,\sigma}$,
 and $\alpha\equiv \rho_{0,+}/\rho_{0,-}$.\\

For simplicity, we make the assumption that spin splitting is large ($\Dels,\Deld \gg \Delc, \dwidth$) and 
$\EF$ lies in the d$\varepsilon$ states of $+$ spin level ($\EF\sim E_{\varepsilon}-\Deld$).
In this configuration, the relation 
\begin{equation} \label{eq:condition}
|g_{m,+}| \gg |g_{m,-}|
\end{equation} 
Holds; then, we can neglect the term including $g_{m,-}$.
Subsequently, $\rho'_{-}(\phi)$ in Eq.\eqref{eq:resistivity1} is neglected; therefore, the AMR ratio in Eq.\eqref{eq:AMRratio} is written as
\begin{align} \label{eq:Perturbation_Result}
\AMR(\phi)&=\frac{\rho_0-\rho_\perp}{\rho_{\perp}}
- \coef\Im\left[ \sum_{m,m'} X^{\ast}_{2,m'}\left(\frac{\pi}{2},0\right)X_{2,m}\left(\frac{\pi}{2},0\right) 
g_{m,+}T_{m,+;m',+}(\phi)g_{m',+}  \right],\\
w&\equiv \frac{15}{4\pi}\frac{1}{(1+\rate)^{2}}\frac{1}{\rho_{\perp}}\frac{m^{\ast}}{e^{2} n_{+}}\frac{\nimp |f(\kF)|^{2}}{\hbar}.
\end{align}
We observe that the finite $c_{4}$ in cubic symmetry can be obtained from the fourth-order perturbation term of SOI (see Appendix) as
\begin{equation}\label{eq:ResultFourfold}
c_{4}\simeq -\frac{\coef}{4} \left(\frac{\lambda}{2}\right)^{4} \Im[g_{\gamma,+}^{2} g_{\varepsilon,+}^{2} 
(4g_{\gamma,+}-g_{\varepsilon,+})].
\end{equation}\\

Next, we discuss how $c_4$ originates as a fourth-order perturbation of SOI in cubic symmetry. 
According to Kokado's study,\cite{Kokado2015}
the $c_4$ term is connected to the difference of the PDOS at the $\EF$ among the d$\varepsilon$-states,
which is realized by the tetragonal distortion in their model.
In the present case, we see that the second-order effect of SOI among the fourth-order perturbation terms of SOI plays a role to split the d$\varepsilon$ states
even in the cubic system.
Then, spin-orbit splitting due to the $\lambda^{2}$ is responsible for the $c_{4}$ term;
In conjunction with the $\lambda^{2}$ contribution that causes conventional AMR, $c_4$ originates the from fourth-order $\lambda$.\\

\subsection{Numerical calculation}
\begin{figure}[H]
	\centering
	\includegraphics[clip, width=0.6\columnwidth]{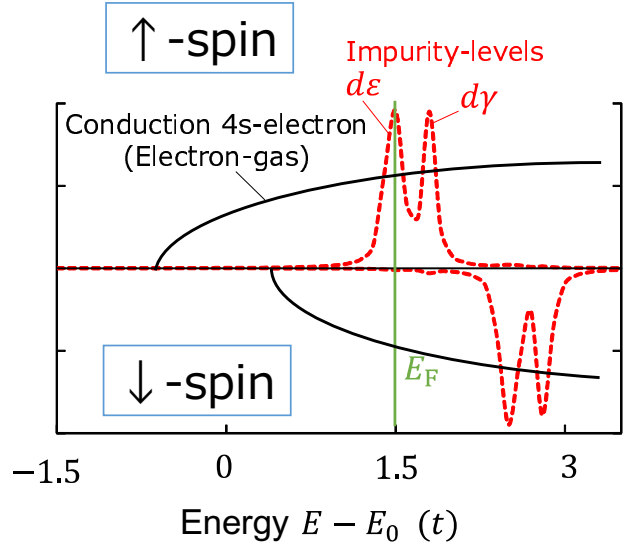}
	\caption{(Color online) PDOS of the conduction states and impurity 3d states projected into $+$ and $-$ spin states.
		The energy levels are measured from $E_{0}$.}
	\label{fig:DOS}
\end{figure}

\begin{figure}[h]
	\begin{minipage}[B]{0.45\columnwidth}
		\centering
		\includegraphics[width=\linewidth]{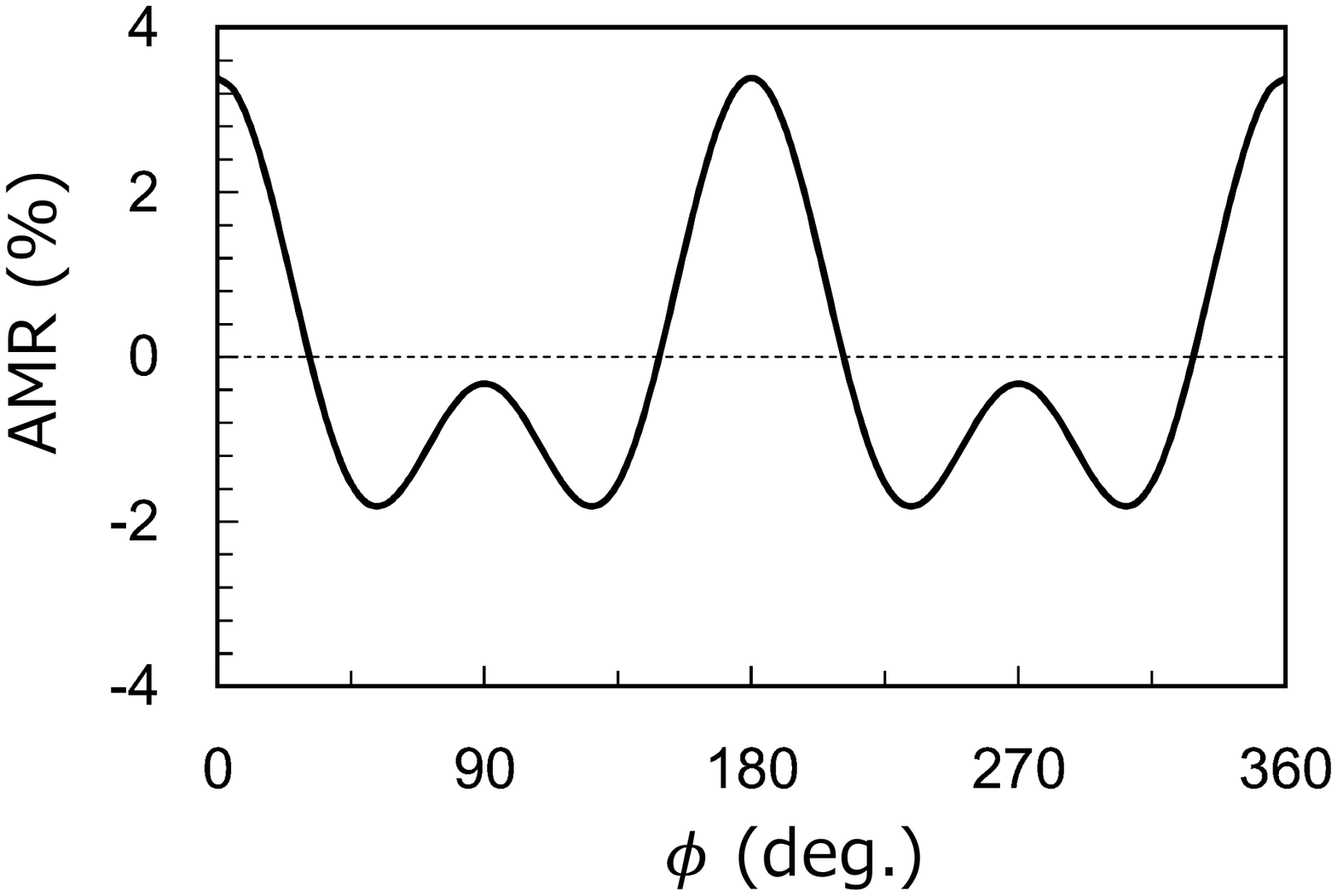}
		\caption{Angular dependence of the AMR ratio with 
			$\Dels=\Deld=0.5, \Delc=0.3, \lambda=0.1, \dwidth=0.05, |f(\kF)|=1.0, \nimp=0.1$, 
			and $\EF\sim E_{\varepsilon}-\Deld$.}
		\label{fig:AMR_deg}
	\end{minipage}
	\hspace{2em}
	\begin{minipage}[B]{0.45\columnwidth}
		\centering
		\includegraphics[width=\linewidth]{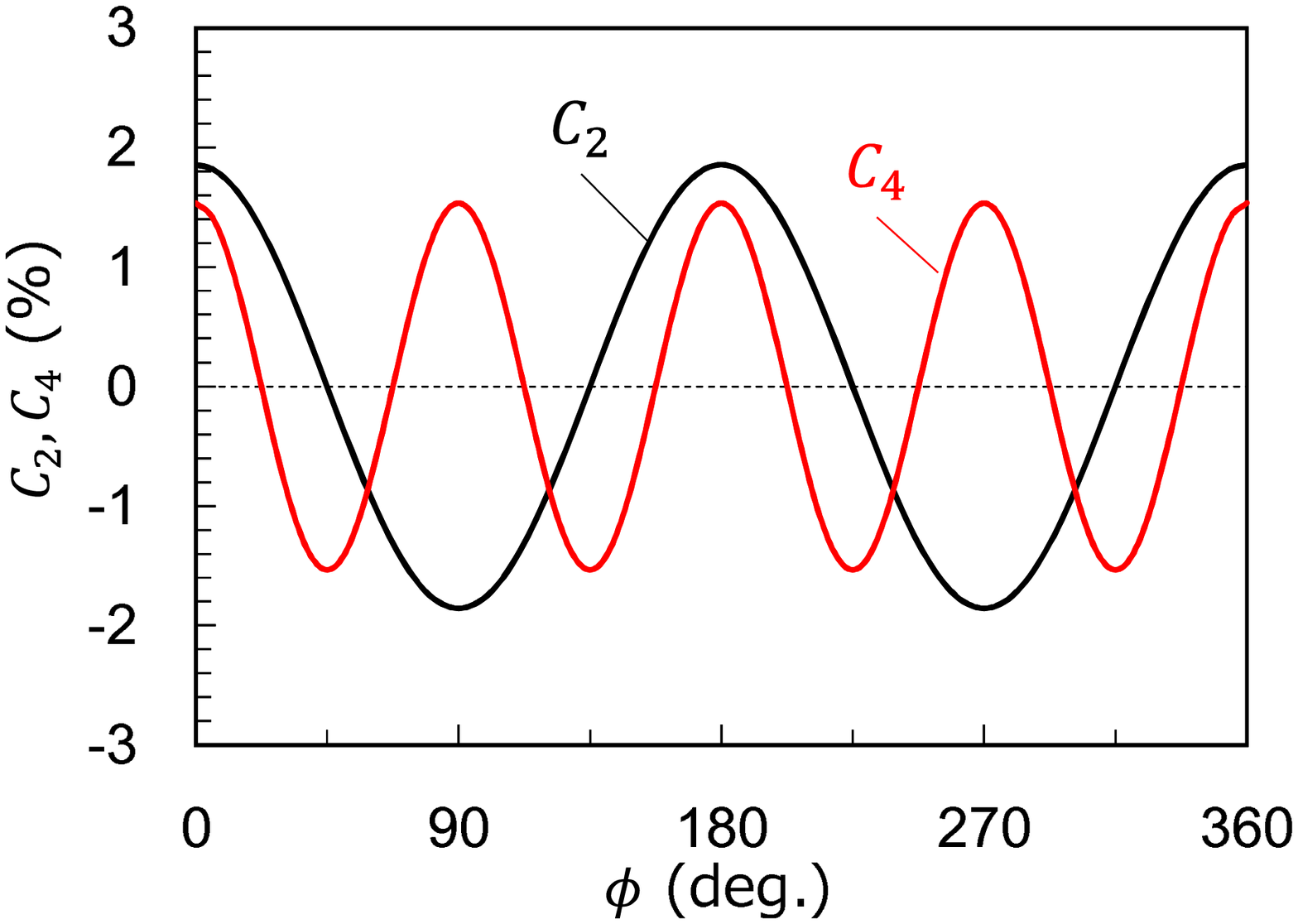}
		\caption{(Color online) Angular dependence of the two-fold and four-fold components of the AMR ratio.}
		\label{fig:decomposedAMR_deg}
	\end{minipage}
\end{figure}

\begin{figure}[H]
	\centering
	\includegraphics[clip, width=0.6\columnwidth]{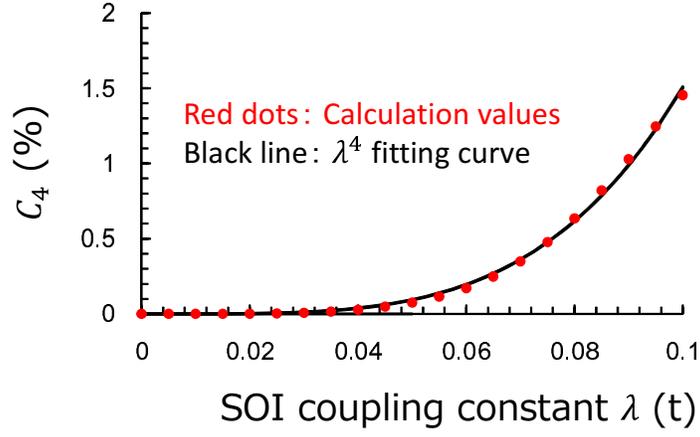}
	\caption{(Color online) The SOI strength dependence of $c_{4}$ and the $\lambda^{4}$ fitting curve,
		with $\Dels=\Deld=0.5, \Delc=0.3, \dwidth=0.05, |f(\kF)|=1.0, \nimp=0.1$, 
		and $\EF\sim E_{\varepsilon}-\Deld$.}
	\label{fig:AMRvsSOI}
\end{figure}

Equation \eqref{eq:KG_2} is directly calculated to qualitatively investigate the AMR behavior. 
As typical values,
we set the parameters $\Dels=\Deld=0.5, \Delc\equiv(E_{\gamma}-E_{\varepsilon})=0.3, \lambda=0.1, \selfenergyss=0.01, \dwidth=0.05, |f(\kF)|=1.0$, and $\nimp=0.1$, in units of $t$. 
Here, the Fermi energy lies in the d$\varepsilon$ level broadened by $\dwidth$ of majority spin bands 
as shown in Fig. \ref{fig:DOS}.
In Fig. \ref{fig:decomposedAMR_deg}, we show the calculated results of AMR, 
and it can be decomposed into the two-fold and four-fold terms. 
Here, it is numerically confirmed that the finite $c_4$ appears even in cubic symmetry. 
Moreover, we calculate the SOI strength $\lambda$ dependence of $c_4$, as shown in Fig.\ref{fig:AMRvsSOI}. 
The intensity of $c_4$ increases with increasing $\lambda$ and is well-fitted by the $\lambda^4$ curve. 
The results indicate that the $c_4$ appears as a fourth-order perturbation effect of SOI, 
supporting the analytical calculation results.\\

\begin{figure}[H]
	\centering
	\includegraphics[width=0.6\linewidth]{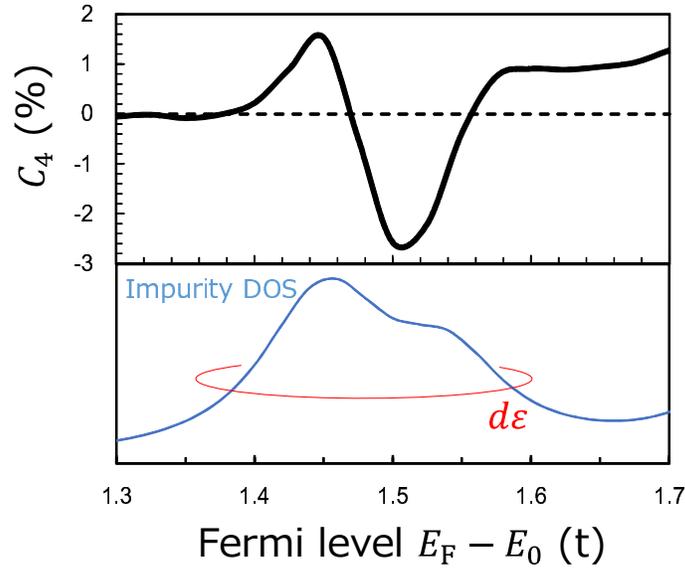}
	\caption{(Color online) $\EF$ dependencies of the $c_{4}$
		and the total DOS of impurity 3d states,
		with $\Dels=\Deld=0.5, \Delc=0.3,  \lambda=0.1, \dwidth=0.05, |f(\kF)|=1.0$, and $\nimp=0.1$.
		In this energy region, the PDOS of d$\varepsilon$ states are dominant in the  impurity DOS. 
	}
	\label{fig:AMRvsFermi}
\end{figure}

Fig.\ref{fig:AMRvsFermi} shows the $\EF$ dependencies of $c_4$ with assuming the impurity's DOS. 
The $c_2$ intensity simply increases when the total DOS of impurity states takes a large value. On the contrary, $c_4$ has a compensation point and can takes both positive and negative values within the same region of ($\EF\sim E_{\varepsilon,+}$). The $c_4$ depends on the properties of each PDOS not the total DOS, hence it is suggested that the $c_4$ has strong material dependence. 
The d-band property should be taken into account to predict the $c_4$s on actual materials.

\section{CONCLUSION}
In summary, we investigate the four-fold AMR ratio in cubic single-crystal ferromagnetic 3d alloys within the s-d scattering model in the presence of the SOI and the cubic symmetric crystal field from a microscopic viewpoint.
The analytical and numerical results indicate that the $c_4$ term appears as a fourth-order perturbation effect of SOI and is sensitive to the PDOSs of 3d states at the Fermi level. 
As a result, we observe that the $c_4$ in the cubic system can be understood as the same scheme as that in the tetragonal system\cite{Kokado2015} by substituting tetragonal splitting with spin-orbit splitting. To explain the material dependence of $c_4$, we need to take into account the material's d-bands structure in a future work.
 
\section*{Acknowledgement}
We would like to thank Professor S. Kokado of Shizuoka University and M. Tsunoda of Tohoku University for their useful discussions.
Yuta YAHAGI acknowledges support from GP-Spin at Tohoku University.
This study was supported by CSRN and JSPS KAKENHI Grant No. 17K14800 in Japan.
We would like to thank Editage (www.editage.com) for English language editing.

\section*{Appendix}
The $c_{4}$ of Eq. \eqref{eq:ResultFourfold} is calculated from the impurity Hamiltonian of Eq. \eqref{eq:HamiltonianImpurity} and the expression of AMR ratio in Eq. \eqref{eq:Perturbation_Result}, 
by taking the perturbation with respect to the SOI in $\Gimp$.
As stated by the degeneracy of $xz$ and $yz$, the calculation is performed by  following a perturbation theory on the degenerate case.

First, in terms of the unperturbed eigenstates , we explicitly write the matrix representation of Hamiltonian.
The unperturbed eigenstates are identified by the combination of 3d orbital levels $m=\varepsilon_{+}, \varepsilon_{0}, \varepsilon_{-},x^{2}-y^{2},3z^{2}-r^{2}$ and spin $\sigma=\pm$ as $\ket{m,\sigma}$.
Here, to avoid difficulty from degeneracy,
we undertake unitary transformation from the subspace of $\{xy,yz,zx\}$ into that of $\{\varepsilon_{+}, \varepsilon_{0}, \varepsilon_{-}\}$ as
\begin{equation}\label{eq:UnitaryTransformation}
\begin{cases}
\ket{\varepsilon_{+},\pm}=\frac{1}{\sqrt{2}}\left\{-\ket{xy,\pm} \pm \i \sin\phi\ket{yz,\pm}\mp \i \cos\phi \ket{xz,\pm} \right\},\\
\ket{\varepsilon_{0},\pm}=\frac{1}{\sqrt{2}}\left\{\cos\phi\ket{yz,\pm}+ \sin\phi \ket{xz,\pm} \right\},\\
\ket{\varepsilon_{-},\pm}=\frac{1}{\sqrt{2}}\left\{\ket{xy,\pm} \pm \i \sin\phi\ket{yz,\pm}\mp \i \cos\phi \ket{xz,\pm} \right\}.
\end{cases}
\end{equation}
The unitary transformation is obtained from the block-diagonalization of $\Hso$ on $\{xy,yz,zx\}$ subspace,
which means that we solve the secular equation in advance.
Therefore, we obtain the unperturbed eigenenergies
\begin{equation}\label{eq:Eigenenergy}
E_{m,\pm}=\begin{cases}
\mp \Deld & m = \varepsilon_{\pm},\varepsilon_{0},\\
\Delc \mp \Deld & m = \dgxy, \dgzr,
\end{cases}
\end{equation}
and matrix represented $\Hso$ as Table. \ref{tab:MatrixTable}.

\begin{table}[htb]
	\centering
	\caption{A part of matrix representation of $\Hso$ (Eq. \eqref{eq:SOI_Hamiltonian}) divided by $\frac{\lambda}{2}$. The remaining elements are obtained from 
		$\braket{m',+|\Hso|m,-}=(\braket{m,-|\Hso|m,+})^{\ast}$ and $\braket{m',-|\Hso|m,-}=(\braket{m',+|\Hso|m,+})^{\ast}$.}
	
	\begin{tabular}{|c|ccccc|}
		\hline 
		$\Hso/\frac{\lambda}{2}$& $\ket{\varepsilon_+,+}$ & $\ket{\varepsilon_0,+}$ & $\ket{\varepsilon_-,+}$ & $\ket{\dgxy,+}$ & $\ket{\dgzr,+}$\\ 
		\hline 
		$\bra{\varepsilon_+,+}$ & 1 & 0 & 0 & 0 & $-\sqrt{\frac{3}{2}} \sin 2\phi$\\ 
		$\bra{\varepsilon_{0},+}$ & 0 & 0 & 0 & $-\i$ & $-\i \sqrt{3} \cos 2\phi$\\ 
		$\bra{\varepsilon_-,+}$ & 0 & 0 & -1 & 0 & $-\sqrt{\frac{3}{2}} \sin 2\phi$\\ 
		$\bra{\dgxy,+}$ & 0 & $\i$ & 0 & 0 & 0 \\ 
		$\bra{\dgzr,+}$ & $-\sqrt{\frac{3}{2}}\sin 2\phi$ & $\i\sqrt{3}\cos 2\phi$ & $-\sqrt{\frac{3}{2}}\sin 2\phi$ & 0 & 0 \\ 
		\hline
		$\bra{\varepsilon_+,-}$ & 0 & $\sqrt{2}$ & 0 &  0 & 0 \\ 
		$\bra{\varepsilon_{0},-}$ & -$\sqrt{2}$ & 0 & 0 & 0 & $\i$ \\ 
		$\bra{\varepsilon_-,-}$ & 0 & 0 & 0 & $ -\i \sqrt{2}$ & 0  \\ 
		$\bra{\dgxy,-}$ & 0 & 0 & $\i \sqrt{2}$ & 0 & 0 \\ 
		$\bra{\dgzr,-}$ & $-\i \sqrt{\frac{3}{2}} \cos 2\phi$ & $\sqrt{3} \sin 2\phi$ & $-\i \sqrt{\frac{3}{2}} \cos 2\phi$ & 0 & 0 \\ 
		\hline 
	\end{tabular} 
	\label{tab:MatrixTable}
\end{table}

Next, we derive the $T(\phi)$ from Eq. \eqref{eq:T-matrix} under the condition of Eq. \eqref{eq:condition}
and substitute the results into Eq. \eqref{eq:Perturbation_Result}.
In particular, we need to calculate $T_{\gamma,+;\gamma',+}(\phi), (\gamma, \gamma' = \dgxy, \dgzr)$ 
because otherwise the terms finally become zero due to $X_{2,m}(\hat{\bm{x}})$ becoming zero.
The odd-order terms in $T_{\gamma,+;\gamma',+}(\phi)$ will be cancelled due to the equivalence of both positive and negative contribution. 
The second-order term $T_{\gamma,+;\gamma',+}^{(2)}(\phi)$ is written as
\begin{equation}\label{eq:SecondOrder}
\begin{cases}
T_{\dgxy,+;\dgxy,+}^{(2)}(\phi)\simeq\left(\frac{\lambda}{2}\right)^{2}g_{\varepsilon,+},\\
T_{\dgxy,+;\dgzr,+}^{(2)}(\phi)=T_{\dgzr,+;\dgxy,+}^{(2)}(\phi)\simeq\sqrt{3}\left(\frac{\lambda}{2}\right)^{2}g_{\varepsilon,+}\cos 2\phi,\\
T_{\dgzr,+;\dgzr,+}^{(2)}(\phi)\simeq 3\left(\frac{\lambda}{2}\right)^{2}g_{\varepsilon,+},
\end{cases}
\end{equation}
where $g_{\varepsilon,+}\equiv(\EF-E_{\varepsilon,+}-\i\dwidth)^{-1}, \varepsilon=\varepsilon_{\pm},\varepsilon_{0}$.
Substituting it into Eq. \eqref{eq:Perturbation_Result}, the resistivity obtains an angular dependence of $\cos2\phi$ and its coefficient can be written as
\begin{equation}
c_{2}\simeq-w\left(\frac{\lambda}{2}\right)^{2}\Im[(g_{\gamma,+})^{2}g_{\varepsilon,+}].
\end{equation}
Therefore, consistent with previous studies, conventional AMR behavior is described by the second-order perturbation theory with respect to the SOI.

The fourth-order term  $T_{\gamma,+;\gamma',+}^{(4)}(\phi)$ is expressed as
\begin{equation}\label{eq:FourthOrder}
\begin{cases}
T_{\dgxy,+;\dgxy,+}^{(2)}(\phi)\simeq\frac{1}{2}\left(\frac{\lambda}{2}\right)^{2}(g_{\varepsilon,+})^{2}g_{\gamma,+}(5+3\cos 4\phi),\\
T_{\dgxy,+;\dgzr,+}^{(2)}(\phi)=T_{\dgzr,+;\dgxy,+}^{(2)}(\phi)\simeq
4\sqrt{3}\left(\frac{\lambda}{2}\right)^{2}(g_{\varepsilon,+})^{2}g_{\gamma,+}\cos 2\phi,\\
T_{\dgzr,+;\dgzr,+}^{(2)}(\phi)\simeq \frac{3}{2}\left(\frac{\lambda}{2}\right)^{2}(g_{\varepsilon,+})^{2}
\left\{(7g_{\gamma,+}+g_{\varepsilon,+})+(g_{\gamma,+}-g_{\varepsilon,+})\cos 4\phi \right\}.
\end{cases}
\end{equation}
Consequently, we obtain the expression of AMR including $\cos 4 \phi$ and its coefficient as
\begin{equation}
c_{4}\simeq -\frac{\coef}{4} \left(\frac{\lambda}{2}\right)^{4} \Im[g_{\gamma,+}^{2} g_{\varepsilon,+}^{2} 
(4g_{\gamma,+}-g_{\varepsilon,+})].
\end{equation}

\bibliography{References.bib}

\newpage
\section*{Figures}
\begin{figure}[H]
	\centering
	\includegraphics[clip, width=0.6\columnwidth]{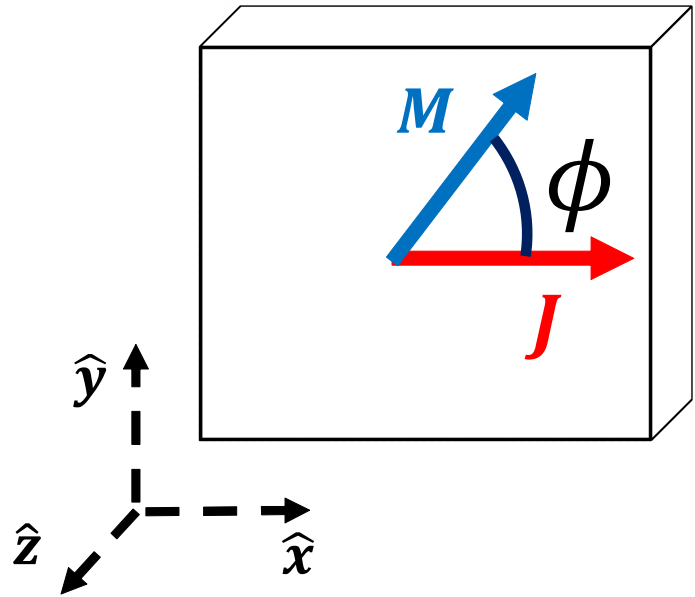}
	\caption{Definition of the angle $\phi$ between the magnetization direction $\hat{\bm{M}}$ and the current direction $\hat{\bm{J}}\parallel \hat{\bm{x}}$. }
	\label{fig:Coordinate}
\end{figure}
\end{document}